\documentclass[bib]{statapress}

\usepackage[crop,newcenter,frame]{pagedims}

\usepackage{sj}

\usepackage{epsfig}

\usepackage{stata}

\usepackage{pdflscape}
\usepackage[flushleft]{threeparttable}
\usepackage{booktabs,caption,fixltx2e}
\usepackage{multirow}

\usepackage{shadow}
\usepackage{amsmath}
\usepackage{amsfonts}
\def\ft#1#2{{\textstyle {\frac{#1}{#2}} }}

\sjsetissue{$vv$}{$ii$}{$mm$}{$yyyy$}

\begin{document}

\inserttype[st0001]{article}


\newcommand{\argmax}{\operatorname*{argmax}}
\newcommand{\argmin}{\operatorname*{argmin}}
\newcommand{\plim}{\operatorname*{plim}}
\newcommand{\E}{\mathbb{E}_{\phi}}
\newcommand{\EE}{\overline{\mathbb{E}}}
\def\ft#1#2{{\textstyle {\frac{#1}{#2}} }}


\author{Cruz-Gonzalez, Fernandez-Val and Weidner}{%
  Mario Cruz-Gonzalez\\Department of Economics\\Boston University\\Boston, MA\\mgonza@bu.edu
  \and
  Iv\'an Fern\'andez-Val\\Department of Economics\\Boston University\\Boston, MA\\ivanf@bu.edu
  \and
  Martin Weidner\\Department of Economics\\University College London\\London, UK\\m.weidner@ucl.ac.uk
}
\title[probitfe and logitfe]{probitfe and logitfe: Bias corrections for probit and logit models with two-way fixed effects}
\maketitle

\begin{abstract}
We present
the Stata commands  \rref{probitfe} and \rref{logitfe}, which  estimate  probit and logit panel data models with individual and/or time unobserved effects.
Fixed effect panel data methods that estimate the unobserved effects can be severely biased because of the incidental parameter problem (Neyman and Scott, \citeyear{Neyman:1948p881}). We tackle this problem by
using the analytical and jackknife bias corrections derived in Fernandez-Val and Weidner (\citeyear{Fernandez-ValandWeidner}) for panels where the two dimensions ($N$ and $T$) are moderately large.
We illustrate the commands with an empirical application to international trade and a Monte Carlo simulation calibrated to this application.

\end{abstract}

\section{Introduction}

Panel data, consisting of multiple observations over time for a set of individuals, are commonly used in empirical analysis to control for unobserved individual and time heterogeneity.  This is often done by adding individual and time effects to the model and treat these unobserved effects as parameter to be estimated in the so-called fixed effect approach. However,  fixed effect estimators of nonlinear  models such as binary response models suffer from the incidental parameter problem (Neyman and Scott, \citeyear{Neyman:1948p881}). 
A special case is the logit model with individual effects where one can use the conditional likelihood approach (Rasch, \citeyear{Rasch1960}; Andersen, 
\citeyear{Andersen1973}; Chamberlain, \citeyear{Chamberlain1984}), implemented in \rref{clogit} and \rref{xtlogit}.  
This approach provides estimates of model coefficients, but is not available for the probit model,
and also does not produce estimates of average partial or marginal effects, which are often the quantities of interest in binary response models. Moreover,  \rref{clogit} and \rref{xtlogit}  do not work well when the panel is long and the model also includes  time effects, because the estimation of the time effects introduces additional incidental parameter bias. Time effects are routinely used in empirical analysis to control for aggregate common shocks and to parsimoniously account for cross sectional dependence.

We deal with the incidental parameter problem by using the bias corrections recently developed by Fernandez-Val and Weidner (\citeyear{Fernandez-ValandWeidner}) for nonlinear panel models with two-way fixed effects. These corrections apply to panel datasets or other pseudo-panel data structures where the two dimensions ($N$ and $T$)  are moderately large; see  Arellano and Hahn (\citeyear{ArellanoHahn2007}) for a survey on bias correction methods to deal with the incidental parameter problem. 
Examples of moderately long panel datasets include traditional microeconomic panel surveys with a long history of data
such as the PSID and NLSY, international cross-country panels such as the PennWorld Table, U.S. state
level panels over time such as the CPS, and square pseudo-panels of trade  flows across countries such
as the Feenstra's World Trade Flows and CEPII, where the indices correspond to the same countries indexed as importers and exporters. The commands \rref{probitfe} and \rref{logitfe} implement analytical and jackknife corrections for fixed effect estimators of  logit and probit models with individual and/or time effects. They produce corrected estimates of the model coefficients and average partial effects.
To the best of our knowledge, these are the first commands in Stata to implement bias correction methods for nonlinear panel models. 

The ado and help files for the commands  \rref{probitfe} and \rref{logitfe} are available through the 
Statistical Software Components of  the Department of Economics of Boston College at \url{http://econpapers.repec.org/software/bocbocode/s458279.htm}  and  \url{http://econpapers.repec.org/software/bocbocode/s458278.htm}.

\paragraph{Notation.} The symbols  $\to_P$ and $\to_d$ are used to denote convergence in probability and distribution, respectively.

\paragraph{Outline.} The rest of the article is organized as follows. Section~2 describes the probit and logit panel models, the incidental parameter problem, and  the bias corrections of  Fernandez-Val and Weidner (\citeyear{Fernandez-ValandWeidner}). Section~3 presents the commands' features. Section~4 provides an illustrative empirical application on international trade flows across countries, together with the results of a Monte Carlo simulation calibrated to the application. We refer the interested reader to Fernandez-Val and Weidner (\citeyear{Fernandez-ValandWeidner}) for details on the assumptions, asymptotic theory and proofs of all the results presented in Section~2.

\section{Probit and Logit Models with Two-Way Fixed Effects}

\subsection{Models and Estimators}

We observe a binary response variable $Y_{it} \in \{0,1\}$ together with a vector of covariates $X_{it}$
for individual $i=1,\ldots,N$ at time  $t=1,\ldots,T$. This definition of the indices $i$ and $t$ applies to standard panel datasets. More generally, $i$ and $t$ can specify any group structure in the data. For example, in the empirical application of Section 4, $i$ and $t$ index the same countries as importers and exporters, respectively.  
 The logit and probit models specify the probability of 
$Y_{it}=1$ conditional on current and past values of the regressors $X^t_i = (X_{i1}, \ldots, X_{it})$, unobserved individual specific effects  $\alpha = (\alpha_1, \ldots, \alpha_N)$, and unobserved time specific effects  $\gamma = (\gamma_1, \ldots, \gamma_T)$, namely
\begin{align*}
 \mathbf{Pr} \left( Y_{it} =1  \mid X^t_i, \alpha, \gamma, \beta \right)
   &= F(X_{it}'\beta+ \alpha_i + \gamma_t),
\end{align*}
where $F: \mathbb{R} \to [0,1]$ is a cumulative distribution function (the standard normal distribution in the probit model and the standard logistic distribution in the logit model), and
 $\beta$ is a vector of unknown model coefficients of the same dimension as $X_{it}$.    The vector $X_{it}$ contains predetermined variables with respect to $Y_{it}$. In particular, $X_{it}$ can include lags of $Y_{it}$ to accommodate dynamic models.  In some static models or in panels where $t$ does not index time,  $X_{it}$ can be treated as strictly exogenous with respect to $Y_{it}$  by replacing $X^t_i$ by $X_i = (X_{i1}, \ldots, X_{iT})$ in the conditioning set. The model does not impose any restriction on the relation between the covariate vector and the unobserved effects. In empirical applications the conditioning on the unobserved effects serves to control for endogeneity as the individual and time effects  capture unobserved heterogeneity that can be related to the covariates.  


We adopt a fixed effect approach and treat the individual and time effects as parameters to be estimated. We denote by $\beta^0$, $\alpha^0$ and $\gamma^0$ the true values of the parameters, that is, the parameters that are assumed to generate the distribution of $Y_{it}$
according the model above. The (conditional) log-likelihood function of the observation $(i,t)$ is
\begin{align*}
     \ell_{it}(\beta, \alpha_i, \gamma_t) &:=  Y_{it} \, \times \log \left[F(X_{it}'\beta+ \alpha_i + \gamma_t)\right]
        + (1-Y_{it}) \, \times \log \left[ 1 - F(X_{it}'\beta+ \alpha_i + \gamma_t)\right],
\end{align*}
and the fixed effect estimators for  $\beta$, $\alpha$ and $\gamma$ are obtained by maximizing the log-likelihood
function of the sample,
\begin{align}\label{eq:fe}
   \left( \widehat \beta, \widehat \alpha, \widehat \gamma \right)
    &\in \argmax_{(\beta,\alpha,\gamma) \in \mathbb{R}^{\dim \beta + N +T}} \sum_{i,t}  \ell_{it}(\beta, \alpha_i, \gamma_t).
\end{align}
This is a smooth concave maximization program for the logit and probit models.
However, there is a perfect collinearity problem because the log-likelihood function is invariant to the transformation
$\alpha_i \mapsto \alpha_i + c$ and $\gamma_t \mapsto \gamma_t - c$ for any $c \in \mathbb{R}$. If $X_{it}$ includes a constant term, we overcome this problem  by dropping $\alpha_1$ and $\gamma_1$, which normalizes $\alpha_1 = 0$ and $\gamma_1 = 0$. If $X_{it}$ does not include a constant term, we only need to  drop either $\alpha_1$ or $\gamma_1$.   
As in linear panel models, the covariates $X_{it}$, other than the constant term, need to vary both across $i$ and over $t$ to avoid further perfect collinearity problems, that is, to guarantee that the log-likelihood function is {\it strictly} concave.

The above fixed effect estimators can be implemented in Stata by using the existing \rref{logit} and \rref{probit}
commands including individual and time binary indicators to account for $\alpha_i$
and $\gamma_t$. However, as we will explain in the next subsection, the fixed effect estimator $\widehat \beta$ can be
severely biased, and the existing routines do not incorporate any bias correction method.

In many applications of the logit and probit models the ultimate parameters of interest are average partial effects (APEs) of the covariates,
which take the form
\begin{equation}
 \delta^0 = \mathbb{E} [\Delta(\beta^0, \alpha^0, \gamma^0)], \ \ \Delta(\beta,  \alpha, \gamma) = (NT)^{-1} \sum_{i,t}   \Delta(X_{it}, \beta, \alpha_i, \gamma_t),
   \label{DefAPE}
\end{equation}
where $\mathbb{E}$ denotes the expectation with respect to the joint distribution of the data and the unobserved  effects.
The expression of the partial effect function $\Delta(X_{it}, \beta, \alpha_i, \gamma_t)$ depends on the type of covariate.
 If $X_{it,k}$, the $k^{\rm th}$ element of $X_{it}$, is binary, then its partial effect on the conditional probability of $Y_{it}$ is
 calculated using
\begin{equation*}
\Delta(X_{it}, \beta , \alpha_i, \gamma_t) =  F(\beta_k +
X_{it,-k}'\beta_{-k} +  \alpha_i + \gamma_t) -
F(X_{it,-k}'\beta_{-k} + \alpha_i +  \gamma_t),
\end{equation*}
where $\beta_k$ is the $k^{\rm th}$ element of $\beta$, and $X_{it,-k}$ and $\beta_{-k}$ include all elements of $X_{it}$ and $\beta$ except for the $k^{\rm th}$ element. This partial effect measures the impact of changing $X_{it,k}$ from $0$ to $1$ on the conditional probability of $Y_{it} = 1$ holding the rest of the covariates fixed at their observed values $X_{it,-k}$. If $X_{it,k}$ is not binary, then the partial effect of $X_{it,k}$ on the conditional probability of $Y_{it}$ is calculated using
\begin{equation*}
\Delta(X_{it}, \beta,\alpha_i, \gamma_t) = \beta_k  \partial F(X_{it}'\beta +
\alpha_i + \gamma_t),
\end{equation*}
where $\partial F$ is the derivative of $F$. This partial effect measures the impact of a marginal change in $X_{it,k}$ on the probability of $Y_{it} =1$ conditional on the observed value of the covariates $X_{it}$.

The fixed effect estimator of  APEs is obtained by  plugging-in estimators of the model parameters in the sample analog of equation
\eqref{DefAPE}, that is,
\begin{equation*}
\widetilde \delta  =  \Delta(\widetilde \beta, \widetilde \alpha, \widetilde \gamma),
\end{equation*}
where $\widetilde \beta$ is an estimator for $\beta$,
and 
\begin{align*}
   \left(\widetilde \alpha, \widetilde \gamma \right)
    &\in \argmax_{(\alpha,\gamma) \in \mathbb{R}^{N +T}} \sum_{i,t}  \ell_{it}(\widetilde \beta, \alpha_i, \gamma_t).
\end{align*}
For example, if $\widetilde \beta = \widehat \beta$ then $(\widetilde \alpha, \widetilde \gamma) = (\widehat \alpha, \widehat \gamma)$, where $(\widehat \beta,\widehat \alpha, \widehat \gamma)$ is the fixed effect estimator defined in \eqref{eq:fe}.
Again, there are Stata routines to calculate $\widetilde \delta$, but they do not implement any bias correction.

\subsection{Incidental Parameter Problem}

The fixed effect estimators $\widehat \beta$ and $\widetilde \delta$ suffer from the Neyman and Scott
incidental parameter problem. In particular, these estimators are  inconsistent under asymptotic 
sequences where $T$ is fixed as $N \to \infty$ when the model has individual effects,  or where $N$ is fixed as $T \to \infty$
when the model has time effects. The source of the problem is that there is only a fixed number of observations to estimate each unobserved effect,  $T$ observations for each individual effect or $N$ observations for each time effect, rendering the corresponding estimators inconsistent. The nonlinearity of the model propagates the inconsistency in the estimation of the individual or time effects to all the model coefficients and APEs.

A recent response to the incidental parameter problem is to consider an alternative asymptotic approximation where $N \to \infty$ and $T \to \infty$ (e.g., Arellano and Hahn, \citeyear{ArellanoHahn2007}). The key insight of this so-called large-$T$ panel data literature is that under this approximation the incidental parameters problem becomes a bias problem that is easier to handle than the inconsistency problem under the traditional asymptotic approximation. 
In particular, Fernandez-Val and Weidner (\citeyear{Fernandez-ValandWeidner}) show that as $N,T \to \infty$, 
with $N/T \to c > 0$, the limit distribution of $ \widehat \beta$ is described by
\begin{align*}
\sqrt{NT}(\widehat \beta - \beta^0 - B^{\beta}/T - D^{\beta}/N)  & \to_d \mathcal{N}(0 ,  V^{\beta}),
\end{align*}
where $ V^{\beta}$ is the asymptotic variance-covariance matrix, $B^{\beta}$ is an asymptotic  bias term coming from the estimation of the individual effects, and  $D^{\beta}$ is an asymptotic  bias term coming from the estimation of the time effects.\footnote{The expressions of $ V^{\beta}$, $B^{\beta}$, and  $D^{\beta}$ for probit and logit models are given in the Appendix.}
 The finite sample prediction of this result is that the fixed effect estimator can have significant bias relative to its dispersion even if $N$ and $T$ are of the same order. Moreover, confidence intervals constructed around the fixed effect estimator can severely undercover the true value of the parameter even in large samples. We show that this large-$N$ large-$T$ version of the incidental parameters problem provides a good approximation to the finite sample behavior of the fixed effect estimator through simulation examples in Section 4.

For $\widetilde \delta$ the situation is different, because the order of  the standard derviation
of $\widetilde \delta$, $1/\sqrt{N} + 1/\sqrt{T}$, is slower than the order of the standard deviation of $\widehat \beta$, $1/\sqrt{NT}$. In this case, Fernandez-Val and Weidner (\citeyear{Fernandez-ValandWeidner}) show that as $N,T \to \infty$, 
with $N/T \to c > 0$, the limit distribution is
\begin{align*}
\sqrt{\min(N,T)}(\widetilde \delta - \delta^0 - B^{\delta}/T - D^{\delta}/N)  & \to_d \mathcal{N}(0 ,  V^{\delta}),
\end{align*}
where $ V^{\delta}$ is the asymptotic variance, $B^{\delta}$ is the asymptotic  bias coming from the estimation of the individual effects, and  $D^{\delta}$ is the asymptotic  bias term coming from the estimation of the time effects.\footnote{The expressions of $ V^{\delta}$, $B^{\delta}$, and  $D^{\delta}$ for probit and logit models are given in the Appendix.}
 Here the standard deviation dominates both of the bias terms, implying that $\widetilde \delta$
is asymptotically first order unbiased. The biases can nevertheless be significant in small
samples as we show in Section 4 through simulation examples.

\subsection{Analytical Bias Correction}
\label{sec:AnalBiasCorr}

The analytical bias correction consists of removing estimates of the leading terms of the bias from the fixed effect estimator of $\beta$. Let $\widehat B^{\beta}$ and $\widehat D^{\beta}$ be consistent
estimators of $ B^{\beta}$ and $ D^{\beta}$, i.e. $\widehat B^{\beta} \to_P B^{\beta}$ and $\widehat D^{\beta} \to_P D^{\beta}$ as $N,T \to \infty$. The bias corrected estimator can be formed as
\begin{equation*}
\widetilde{\beta}^A = \widehat{\beta} -  \widehat{B}^{\beta}/T - \widehat D^{\beta}/N.
\end{equation*}
As $N, T \to \infty$ with $N/T \to c > 0$ the limit distribution of $\widetilde{\beta}^A$ is
\begin{equation*}
\sqrt{NT}(\widetilde \beta^A - \beta^0)
\to_d \mathcal{N}(0,  V^{\beta}).
\end{equation*}
The analytical correction therefore centers the asymptotic distribution at the true value of the parameter, without increasing asymptotic variance.
This  result predicts that in large samples the corrected estimator has small bias relative to dispersion, the correction does not increase dispersion, and the confidence intervals constructed around the corrected estimator have coverage probabilities close to the nominal levels. We show that these predictions provide a good approximation to the behavior of the corrections in Section 4.

The bias corrected APEs can be constructed in the same fashion as
\begin{equation*}
\widetilde{\delta}^A = \widetilde \delta -  \widehat{B}^{\delta}/T - \widehat D^{\delta}/N,
\end{equation*}
where $\widehat B^{\delta}$ and $\widehat D^{\delta}$ are consistent estimators 
of $ B^{\delta}$ and $ D^{\delta}$, i.e. $\widehat B^{\delta} \to_P B^{\delta}$ and $\widehat D^{\delta} \to_P D^{\delta}$
as $N, T \to \infty$.  The limit distribution of $\widetilde{\delta}^A$ is
\begin{equation*}
\sqrt{\min(N,T)}(\widetilde \delta^A - \delta^0  + o_P(T^{-1} + N^{-1}))
\to_d \mathcal{N}(0,  V^{\delta}).
\end{equation*}

We give the details on how to compute $\widehat B^{\beta}$, $\widehat D^{\beta}$, $\widehat{B}^{\delta}$
and $\widehat D^{\delta}$  in the Appendix.
The {\tt probitfe} and {\tt logitfe} commands compute these analytical bias corrections with the
option  {\tt \underbar{an}alytical}. When the regressors $X_{it}$ are predetermined, e.g.~when lagged dependent
variables are included, then the calculation of $\widehat B^{\beta}$ 
and $\widehat{B}^{\delta}$, and thus of the bias corrections, requires the specification of a trimming 
parameter $L \in \{1,2,3,\ldots\}$ in order to estimate a spectral expectation. For the asymptotic theory the requirement 
on $L$ is that  $L \to \infty$ such that $L/T \to 0$. 
In practice, we do not recommend to use $L$ larger than four, and we suggest
to compute the analytical bias corrections for different values of $L$ as a robustness check. 
 When the regressors $X_{it}$ are strictly exogenous, then $L$ 
should be set to zero. The trimming parameter is set through the command options {\tt \underbar{lags}({\it integer})},
as described below.

\subsection{Jackknife Bias Correction}

The commands {\tt probitfe} and {\tt logitfe} with the {\tt \dunderbar{jack}knife} option allow for six different types of jackknife corrections, denoted as {\tt ss1}, {\tt ss2}, {\tt js}, {\tt sj}, {\tt jj} and {\tt double}.  We will briefly explain each correction and give some intuition about how they reduce bias. 
%
%
The jackknife corrections do not require explicit estimation of the bias, but are computationally more intensive as they involve solving multiple fixed effect estimation programs. 
The methods 
%
%
are combinations of the leave-one-observation-out panel jackknife (PJ) of Hahn and Newey (\citeyear{Hahn:2004p882}) and the split panel jackknife (SPJ) of Dhaene and Jochmans (\citeyear{DhaeneJochmans2015}) applied to the two dimensions of the panel.
 
 Let ${\bf N} =  \{ 1,\ldots, N\}$ and ${\bf T} =  \{ 1,\ldots, T\}$.
Define the fixed effect estimator of $\beta$ in the subpanel with cross sectional indices 
$A \subseteq {\bf N}$ and time series indices $B \subseteq {\bf T}$ as
$$
\widehat \beta_{A,B} \in  \argmax_{\beta \in \mathbb{R}^{\dim \beta} } \;
   \max_{\alpha(A) \in \mathbb{R}^{|A|}}  \;
    \max_{\gamma(B) \in \mathbb{R}^{|B|}} 
\; \sum_{i \in A,t \in B}   \;  \ell_{it}(\beta,\alpha_i,\gamma_t)  ,
$$
where $\alpha(A)=\{ \alpha_i  :  i \in A \}$ and 
$\gamma(B)=\{ \gamma_t   :   t \in B \}$.
Notice that the original fixed effect estimator $\widehat \beta$ defined above is 
equal to $\widehat \beta_{{\bf N},{\bf T}}$.
Using this notation we can now describe the six jackknife corrections:

\begin{itemize}
\item
The correction {\tt ss1} applies simultaneously SPJ to both dimensions of the panel. Let $\widetilde{\beta}_{N/2,T/2}$ be the average of the four split jackknife estimators that leave out half of the individuals and half of the time periods, that is
\begin{align*}
    \widetilde{\beta}_{N/2,T/2}
      &= \frac 1 4 
      \bigg[ 
              \widehat \beta_{\{i \, : \, i \leq \lceil N/2 \rceil \}, \{t \, : \, t \leq \lceil T/2 \rceil \}}
         +    \widehat \beta_{\{i \, : \, i \geq \lfloor N/2+1 \rfloor \}, \{t \, : \, t \leq \lceil T/2 \rceil \}}
       \\ & \qquad   \; \;
         +   \widehat \beta_{\{i \, : \, i \leq \lceil N/2 \rceil \}, \{t \, : \, t  \geq \lfloor T/2+1 \rfloor \}}
         +    \widehat \beta_{\{i \, : \, i \geq \lfloor N/2+ \rfloor \}, \{t \, : \, t  \geq \lfloor T/2+1  \rfloor \}}
      \bigg]  ,
\end{align*}
where $\lfloor . \rfloor$ and $\lceil . \rceil$ denote the floor and ceiling function, respectively.
The {\tt ss1} corrected estimator is 
\begin{equation*}
\widetilde{\beta}^{\rm ss1}=2\widehat{\beta} - \widetilde{\beta}_{N/2,T/2} .
\end{equation*}

\item
The correction {\tt ss2} applies separately SPJ to both dimensions of the panel. Let $\widetilde{\beta}_{N,T/2}$ be the average of the two split jackknife estimators that leave out the first and second halves of the time periods, and let $\widetilde{\beta}_{N/2,T}$ be the average of the two split jackknife estimators that leave out half of the individuals, that is,
\begin{align*}
      \widetilde{\beta}_{N,T/2}
       &= \frac 1 2 
       \left[  \widehat \beta_{{\bf N}, \{t \, : \, t \leq \lceil T/2 \rceil \}}
            +   \widehat \beta_{{\bf N}, \{t \, : \, t  \geq \lfloor T/2+1  \rfloor \}}
       \right] ,
     \\
      \widetilde{\beta}_{N/2,T}
       &=   \frac 1 2 
       \left[       \widehat \beta_{\{i \, : \, i \leq \lceil N/2 \rceil \}, {\bf T}}
         +    \widehat \beta_{\{i \, : \, i \geq \lfloor N/2+1 \rfloor \}, {\bf T}}
       \right]  .
\end{align*}
The {\tt ss2} corrected estimator is
\begin{equation*}
\widetilde{\beta}^{\rm ss2} = 3  \widehat \beta - \widetilde{\beta}_{N,T/2} - \widetilde{\beta}_{N/2,T}.
\end{equation*}

\item
The correction {\tt js} applies PJ to the individual dimension and SPJ to the time dimension. Let
$\widetilde{\beta}_{N,T/2}$ be defined as above,
and let $\widetilde{\beta}_{N-1,T}$ be the average of the $N$ jackknife estimators that leave out one individual, that is,
\begin{align*}
     \widetilde{\beta}_{N-1,T}
       &= \frac 1 N \sum_{i=1}^N 
           \,  \widehat \beta_{{\bf N} \setminus \{i\}, {\bf T}} .
\end{align*}
The {\tt js} corrected estimator is
\begin{equation*}
\widetilde{\beta}^{\rm js} = (N+1)\widehat{\beta} - (N-1)\widetilde{\beta}_{N-1,T} - \widetilde{\beta}_{N,T/2}.
\end{equation*}

\item
The correction {\tt sj} applies SPJ to the individual dimension and PJ to the time dimension. Let
$\widetilde{\beta}_{N/2,T}$ be defined as above,
and let $\widetilde{\beta}_{N,T-1}$ be the average of the $T$ jackknife estimators that leave out one time period, that is,
\begin{align*}
     \widetilde{\beta}_{N,T-1}
       &= \frac 1 T \sum_{t=1}^T 
           \,  \widehat \beta_{{\bf N}, {\bf T} \setminus \{t\}} .
\end{align*}
The {\tt sj} corrected estimator is
\begin{equation*}
\widetilde{\beta}^{\rm sj} = (T+1)\widehat{\beta} - \widetilde{\beta}_{N/2,T} - (T-1)\widetilde{\beta}_{N,T-1}.
\end{equation*}

\item The correction {\tt jj} applies PJ to both the individual and the time dimension. Let $\widetilde{\beta}_{N-1,T}$ 
and $\widetilde{\beta}_{N,T-1}$ be defined as above.
The {\tt jj} corrected estimator is
\begin{equation*}
\widetilde{\beta}^{\rm jj} = (N+T-1)\widehat{\beta} - (N-1)\widetilde{\beta}_{N-1,T} - (T-1)\widetilde{\beta}_{N,T-1}. 
\end{equation*}

\item The correction {\tt double} uses PJ for observations with the same cross section and time series indices. This type of correction only makes sense for panels where $i$ and $t$ index the same entities. For example, in country trade data, the cross-section dimension represents each country as an importer, and the ``time-series dimension'' represents each country as an exporter. Thus, let $N=T$ and define
  $\widetilde{\beta}_{N-1,N-1}$ as the average of the $N$ jackknife estimators that leave one entity out,
that is
\begin{align*}
     \widetilde{\beta}_{N-1,N-1}
       &= \frac 1 N \sum_{i=1}^N 
           \,  \widehat \beta_{{\bf N} \setminus \{i\} , {\bf N} \setminus \{i\}} .
\end{align*}
 The corrected estimator is 
\begin{equation*}
\widetilde{\beta}^{\rm double} = N\widehat{\beta} - (N-1)\widetilde{\beta}_{N-1,N-1}.
\end{equation*}

\end{itemize}

To give some intuition on how these corrections reduce bias, we use a first order approximation to the bias
$$
\text{bias}[\widehat \beta_{A,B}] \approx \widehat{B}^{\beta}/|A| + \widehat D^{\beta}/|B|,
$$
where $|A|$ denotes the cardinality of the set $A$.
Consider for example the option {\tt ss1}. Using the previous approximation
\begin{equation*}
\text{bias}[\widetilde{\beta}^{\rm ss1}]   \approx 2 \times \text{bias}[\widehat{\beta}_{{\bf N},{\bf T}}] - \text{bias}[\widetilde{\beta}_{N/2,T/2}] = 0,
\end{equation*}
because the leading bias of $\widetilde{\beta}_{N/2,T/2}$   is twice the leading bias 
in $\widehat{\beta}_{{\bf N},{\bf T}}$ since  the subpanels used to construct $\widetilde{\beta}_{N/2,T/2}$ contain half of the individuals and time periods.
 In other words, subtracting 
$(\widetilde{\beta}_{N/2,T/2} - \widehat{\beta}_{{\bf N},{\bf T}})$ from $\widehat{\beta}$ removes a nonparametric estimator of the leading bias.
Similarly, we can show   that the leading bias 
of $\widehat{\beta}$ is removed by the other corrections as they use appropriate choices
of the size of the subpanels and corresponding coefficients in the linear combinations of the subpanel estimators.

There are  panels for which there is no natural ordering of the observations along some of the dimensions, e.g. the individuals in the PSID. In this case there are multiple ways to select the subpanels to implement the {\tt ss1} and {\tt ss2} corrections. To avoid any arbitrariness in the choice of subpanels,  the command includes the possibility 
of constructing $\widetilde{\beta}_{N/2,T/2}$ and $\widetilde{\beta}_{N/2,T}$ 
as the average of the estimators obtained from multiple orderings of the panels by randomly permuting the indices of the dimension that has no natural ordering of the observations. The option {\tt \underbar{mul}tiple({\it integer})} allows the user to specify the number of different permutations of the panel to use. 

Fernandez-Val and Weidner (\citeyear{Fernandez-ValandWeidner}) show that the correction {\tt ss2}  removes the bias without increasing  dispersion in large samples. In particular, it is shown that the limit distribution of $\widetilde{\beta}^{\rm ss2}$ as $N,T \to \infty$ with $N/T \to c>0$ is
$$
\sqrt{NT}(\widetilde \beta^{\rm ss2} - \beta^0)
\to_d \mathcal{N}(0,  V^{\beta}),
$$
the same as the limit distribution of the analytical correction. The assumptions required  for this result include  homogeneity conditions along the two dimensions of the panel to guarantee  that the bias terms   $B^{\beta}$ and $  D^{\beta}$ are the same in all the subpanels. The analytical corrections described
 above do not require this type of conditions and are therefore more widely applicable.

%

Jackknife corrections for the APEs are formed analogously.  We
compute estimates $\widetilde \delta_{A,B}$ from subpanels with cross sectional indices $A \subseteq {\bf N}$ and time series indices $B \subseteq {\bf T}$, and use the corrections described above replacing  $\beta$ by $\delta$ everywhere. 

\subsection{One-Way Fixed Effects}

So far we have focused on two-way fixed effect models with individual and time effects because they are the most commonly used  in empirical applications. For completeness, the commands \rref{probitfe} and \rref{logitfe} also provide functionality for one-way fixed effect models that include only either individual effects or time effects (using the options {\bf \tt \underline{ieffects}} and {\bf \tt \underline{teffects}}, respectively), as well as the flexibility to choose whether the bias corrections should account only either for individual effects or time effects (using the options {\bf \tt \underline{ibias}} and {\bf \tt \underline{tbias}}, respectively).
%
%
%
Fixed effect estimators of these models also suffer from the incidental parameter problem.  The commands implement the analytical and jackknife corrections of Hahn and Newey (\citeyear{Hahn:2004p882}) and Fernandez-Val (\citeyear{FernandezVal:2009p3313}), and the split-panel correction of Dhaene and Jochmans  (\citeyear{DhaeneJochmans2015}). We do not describe these corrections in detail, because they are very similar to the ones described above for two-way models. For example, the analytical correction for $\beta$ has the same form as $\widetilde{\beta}^A$ after making one of the estimated bias terms equal to zero: $\widehat D^{\beta} = 0$ for models without time effects, and  $\widehat B^{\beta} = 0$ for models without individual effects. We give the expressions of $B^{\beta}$ and $D^{\beta}$ and describe the jackknife corrections for one-way fixed effect models in the Appendix.

\subsection{Unbalanced Panel Data}

In the description of the incidental parameter problem and bias corrections, we implicitly assumed that the panel was balanced, i.e., we observe
each individual, $i=1,\ldots,N$, at each time period, $t=1,\ldots,T$.  Nevertheless, unbalanced panel datasets are common in empirical applications. 
Unbalancedness does not introduce special theoretical complications provided that the source of the missing observations is random. It does not
introduce complications in the computation either, because \rref{probitfe} and \rref{logitfe} make use of Stata's time-series operators that account for missing observations, provided the data are
declared to be time series.

Suppose, for example, that we have the following dataset:
\begin{stlog}
. tsset 
       panel variable:  id (weakly balanced)
        time variable:  time, 1 to 7, but with gaps
                delta:  1 unit
\end{stlog}

\begin{stlog}
. list id time
{\smallskip}
     {\TLC}\HLI{11}{\TRC}
     {\VBAR} id   time {\VBAR}
     {\LFTT}\HLI{11}{\RGTT}
  1. {\VBAR}  1      1 {\VBAR}
  2. {\VBAR}  1      2 {\VBAR}
  3. {\VBAR}  1      4 {\VBAR}
  4. {\VBAR}  1      5 {\VBAR}
  5. {\VBAR}  1      7 {\VBAR}
     {\LFTT}\HLI{11}{\RGTT}
  6. {\VBAR}  2      2 {\VBAR}
  7. {\VBAR}  2      3 {\VBAR}
  8. {\VBAR}  2      5 {\VBAR}
  9. {\VBAR}  2      6 {\VBAR}
 10. {\VBAR}  2      7 {\VBAR}
     {\BLC}\HLI{11}{\BRC}
\end{stlog}
which includes two individuals and seven time periods, but there is no observations for every time period for each individual. Time-series operators are important when the analytical correction is applied and the trimming parameter is higher than zero. If the trimming parameter is equal to one, for example, \rref{probitfe} and \rref{logitfe} will correctly produce a missing value for $t=\{1,4,7\}$ for the first individual, and for $t=\{2,5\}$ for the second individual.

In the jackknife corrections, \rref{probitfe} and \rref{logitfe}  identify the appropriate subset of observations for each individual, because they use time as index instead of the observation number. If we apply, for example, the jackknife bias correction {\tt ss1}, where the subpanels include half of the time periods for each individual, the commands will correctly use $t=\{1,2,4\}$ for the first individual, and $t=\{2,3\}$ for the second individual.

\section{The probitfe and logitfe commands}

\subsection{Syntax}

Both {\tt probitfe} and {\tt logitfe} share the same syntax and options. We use here the syntax for {\tt probitfe}. The user only needs to replace {\tt logitfe} in place of {\tt probitfe} if she wishes to fit a logit model.

\hspace{-.28in} {\bf Uncorrected (NC) estimator}

\begin{stsyntax}
probitfe
    \depvar\
    {\it indepvars}
    \optif\
    \optin\
    \optional{,
    \underbar{noc}orrection
    NC\_options}
\end{stsyntax}

\hspace{-.28in} {\bf Analytical-corrected (AC) estimator}

\begin{stsyntax}
probitfe
    \depvar\
    {\it indepvars}
    \optif\
    \optin\
    \optional{,
    \underbar{an}alytical
    AC\_options}
\end{stsyntax}

\hspace{-.28in} {\bf Jackknife-corrected (JC) estimator}

\begin{stsyntax}
probitfe
    \depvar\
    {\it indepvars}
    \optif\
    \optin\
    \optional{,
    \underbar{jack}knife
    JC\_options}
\end{stsyntax}

Both, a panel variable and a time variable must be specified. {\it indepvars} may contain factor variables. {\it depvar} and {\it indepvars} may contain time-series operators.

\subsection{Options for uncorrected (NC) estimator}

\subsubsection{Type of Included Effects}

\hangpara
{\tt \underbar{ieffects}(\ststring)} specifies whether the uncorrected estimator includes individual effects.

\morehang
{\tt \underbar{ieffects}(yes)}, the default, includes individual fixed-effects.

\morehang
{\tt \underbar{ieffects}(no)} omits the individual fixed-effects.

\hangpara
{\tt \underbar{teffects}(\ststring)} specifies whether the uncorrected estimator includes time effects.

\morehang
{\tt \underbar{teffects}(yes)}, the default, includes time fixed-effects.

\morehang
{\tt \underbar{teffects}(no)} omits the time fixed-effects.\footnote{{\tt \underbar{ieffects}(no)} and {\tt \underbar{teffects}(no)} is an invalid option.}

\subsubsection{Finite Population Correction}

\hangpara
{\tt \dunderbar{pop}ulation({\it integer})} adjusts the estimation of the variance of the APEs by a finite population correction. Let $m$ be the number of original observations included in {\tt probitfe}, and $M\geq m$ the number of observations for the entire population declared by the user. The computation of the variance of the APEs is corrected by the factor $fpc=(M-m)/(M-1)$. The default is $fpc=1$, corresponding to an infinite population. Notice that $M$ makes reference to the total number of observations and not the total number of individuals. If, for example, the population has 100 individuals followed over 10 time periods, the user must use {\tt \dunderbar{pop}ulation(1000)} instead of {\tt \dunderbar{pop}ulation(100)}.

\subsection{Options for analytical-corrected (AC) estimator}

\subsubsection{Trimming Parameter}

\hangpara
{\tt \underbar{lags}({\it integer})} specifies the value of the trimming parameter to estimate spectral expectations, see 
the discussion in the Section~\ref{sec:AnalBiasCorr} for details. The default is {\tt \underbar{lags}(0)}, that is,
the trimming parameter to estimate spectral expectations is set to zero. This option should be used when the model is static with strictly exogenous regressors.

\morehang
The trimming parameter can be set to any value between zero and $(T-1)$. A trimming parameter higher than zero 
should be used when the model is dynamic or some of the regressors are weakly exogenous or predetermined. As mentioned
in Section~\ref{sec:AnalBiasCorr}, we do not recommend to set the value of the trimming parameter to a value higher than four. 

\subsubsection{Type of Included Effects}

\hangpara
{\tt \underbar{ieffects}(\ststring)} specifies whether the model includes individual fixed-effects.

\morehang
{\tt \underbar{ieffects}(yes)}, the default, includes individual fixed-effects.

\morehang
{\tt \underbar{ieffects}(no)} omits the individual fixed-effects.

\hangpara
{\tt \underbar{teffects}(\ststring)} specifies whether the model includes time fixed-effects.

\morehang
{\tt \underbar{teffects}(yes)}, the default, includes time fixed-effects.

\morehang
{\tt \underbar{teffects}(no)} omits the time fixed-effects.\footnote{{\tt \underbar{ieffects}(no)} and {\tt \underbar{teffects}(no)} is an invalid option.}

\subsubsection{Type of Correction}

\hangpara
{\tt \underbar{ibias}(\ststring)} specifies whether the analytical correction accounts for individual effects.

\morehang
{\tt \underbar{ibias}(yes)}, the default, corrects for the bias coming from the individual fixed-effects.

\morehang
{\tt \underbar{ibias}(no)} omits the individual fixed-effects analytical bias correction.

\hangpara
{\tt \underbar{tbias}(\ststring)} specifies whether the analytical correction accounts for time effects.

\morehang
{\tt \underbar{tbias}(yes)}, the default, corrects for the bias coming from the time fixed-effects.

\morehang
{\tt \underbar{tbias}(no)} omits the time fixed-effects analytical bias correction.\footnote{{\tt \underbar{ibias}(no)} and {\tt \underbar{tbias}(no)} is an invalid option.}

\subsubsection{Finite Population Correction}

\hangpara
{\tt \dunderbar{pop}ulation({\it integer})} adjusts the estimation of the variance of the APEs by a finite population correction. Let $m$ be the number of original observations included in {\tt probitfe}, and $M\geq m$ the number of observations for the entire population declared by the user. The computation of the variance of the APEs is corrected by the factor $fpc=(M-m)/(M-1)$. The default is $fpc=1$, corresponding to an infinite population. Notice that $M$ makes reference to the total number of observations and not the total number of individuals. If, for example, the population has 100 individuals followed over 10 time periods, the user must use {\tt \dunderbar{pop}ulation(1000)} instead of {\tt \dunderbar{pop}ulation(100)}.

\subsection{Options for jackknife-corrected (JC) estimator}

\subsubsection{\# of Partitions}

\hangpara
{\tt ss1} specifies split panel jackknife in four non-overlapping sub-panels; in each sub-panel half of the individuals and half of the time periods are left out. See previous section for the details.

\morehang
{\tt \underbar{mul}tiple({\it integer})} is a {\tt ss1} sub-option that allows for different multiple partitions, each one made on a randomization of the observations in the panel; the default is zero, i.e. the partitions are made on the
original order in the data set. If {\tt \underbar{mul}tiple(10)} is specified, for example, then the {\tt ss1} estimator is computed 10 times on 10 different randomizations of the observations in the panel; the resulting estimator is the mean of these 10 split panel jackknife corrections. This option can be used if there is a dimension of the panel where there is no
natural ordering of the observations.

\morehang
{\tt \underbar{i}ndividuals} specifies the multiple partitions to be made only on the cross-sectional dimension.

\morehang
{\tt \underbar{t}ime} specifies the multiple partitions to be made only on the time dimension.

\morehang
If neither {\tt \underbar{i}ndividuals} nor {\tt \underbar{t}ime} options are specified, the multiple partitions are made on both the cross-sectional and the time dimensions.

\hangpara
{\tt ss2}, the default, specifies split jackknife in both dimensions. As in {\tt ss1}, there are four sub-panels: in two of them half of the individuals are left out but all time periods are included; in the other two half of the time periods are left out but all the individuals are included. See previous section for the details.

\morehang
{\tt \underbar{mul}tiple({\it integer})} is a {\tt ss2} sub-option that allows for different multiple partitions, each one made on a randomization of the observations in the panel; the default is zero, i.e. the partitions are made on the
original order in the data set. If {\tt \underbar{mul}tiple(10)} is specified, for example, then the {\tt ss2} estimator is computed 10 times on 10 different randomizations of the observations in the panel; the resulting estimator is the mean of these 10 split panel jackknife corrections. This option can be used if there is a dimension of the panel where there is no
natural ordering of the observations.

\morehang
{\tt \underbar{i}ndividuals} specifies the multiple partitions to be made only on the cross-sectional dimension.

\morehang
{\tt \underbar{t}ime} specifies the multiple partitions to be made only on the time dimension.

\morehang
If neither {\tt \underbar{i}ndividuals} nor {\tt \underbar{t}ime} options are specified, the multiple partitions are made on both the cross-sectional and the time dimensions.

\hangpara
{\tt js} uses delete-one panel jackknife in the cross-section and split panel jackknife in the time series. See previous section for the details.

\hangpara
{\tt sj} uses split panel jackknife in the cross-section and delete-one jackknife in the time series. See previous section for the details.

\hangpara
{\tt jj} uses delete-one jackknife in both the cross-section and the time series. See previous section for the details.

\hangpara
{\tt double} uses delete-one jackknife for observations with the same cross-section and the time-series indices. See previous section for the details.

\subsubsection{Type of Included Effects}

\hangpara
{\tt \underbar{ieffects}(\ststring)} specifies whether the model includes individual fixed-effects.

\morehang
{\tt \underbar{ieffects}(yes)}, the default, includes individual fixed-effects.

\morehang
{\tt \underbar{ieffects}(no)} omits the individual fixed-effects.

\hangpara
{\tt \underbar{teffects}(\ststring)} specifies whether the model includes time fixed-effects.

\morehang
{\tt \underbar{teffects}(yes)}, the default, includes time fixed-effects.

\morehang
{\tt \underbar{teffects}(no)} omits the time fixed-effects.\footnote{{\tt \underbar{ieffects}(no)} and {\tt \underbar{teffects}(no)} is an invalid option.}

\subsubsection{Type of Correction}

\hangpara
{\tt \underbar{ibias}(\ststring)} specifies whether the jackknife correction accounts for the individual effects.

\morehang
{\tt \underbar{ibias}(yes)}, the default, corrects for the bias coming from the individual fixed-effects.

\morehang
{\tt \underbar{ibias}(no)} omits the individual fixed-effects jackknife correction. If this option and multiple partitions only in the time-dimension are specified togeteher (for the jackknife {\tt ss1/ss2} corrections), the resulting estimator is equivalent to the one without multiple partitions.

\hangpara
{\tt \underbar{tbias}(\ststring)} specifies whether the jackknife correction accounts for the time effects.

\morehang
{\tt \underbar{tbias}(yes)}, the default, corrects for the bias coming from the time fixed-effects.

\morehang
{\tt \underbar{tbias}(no)} omits the time fixed-effects jackknife correction. If this option and multiple partitions only in the cross-section are specified togeteher (for the jackknife {\tt ss1/ss2} corrections), the resulting estimator is equivalent to the one without multiple partitions\footnote{{\tt \underbar{ibias}(no)} and {\tt \underbar{tbias}(no)} is an invalid option.}.

\subsubsection{Finite Population Correction}

\hangpara
{\tt \dunderbar{pop}ulation({\it integer})} adjusts the estimation of the variance of the APEs by a finite population correction. Let $m$ be the number of original observations included in {\tt probitfe}, and $M\geq m$ the number of observations for the entire population declared by the user. The computation of the variance of the APEs is corrected by the factor $fpc=(M-m)/(M-1)$. The default is $fpc=1$, corresponding to an infinite population. Notice that $M$ makes reference to the total number of observations and not the total number of individuals. If, for example, the population has 100 individuals followed over 10 time periods, the user must use {\tt \dunderbar{pop}ulation(1000)} instead of {\tt \dunderbar{pop}ulation(100)}.

\subsection{Saved results}

{\tt probitfe} and {\tt logitfe} save the following in {\tt e()}:

\begin{stresults}
\stresultsgroup{Scalars} \\
\stcmd{e(N)}				& number of observations			& \stcmd{e(N\_drop)}				& number of observations dropped		\\
\stcmd{e(N\_group\_drop)}	& \quad number of groups dropped	&							& \quad because of all positive or all		\\
						& \quad because of  all positive		&							& \quad zero outcomes			\\
						& \quad or all zero outcomes	&							&								\\
\stcmd{e(N\_time\_drop)}		& number of time periods			& \stcmd{e(N\_group)}			& number of groups					\\
						& \quad dropped because of all		& \stcmd{e(T\_min)}				& smallest group size				\\ 
						& \quad positive or all zero		& \stcmd{e(T\_avg)}				& average group size				\\
						& \quad outcomes				& \stcmd{e(T\_max)}				& largest group size					\\
\stcmd{e(k)}				& number of parameters			& \stcmd{e(df\_m)}				& model degrees of freedom			\\ 
						& \quad excluding individual		& \stcmd{e(r2\_p)}				& pseudo R-squared					\\
						& \quad and/or time effects		& \stcmd{e(chi2)}				& likelihood-ratio chi-squared			\\
\stcmd{e(p)}				& significance of model test		&							& \quad model test					\\
\stcmd{e(rankV)}			& rank of \stcmd{e(V)}			& \stcmd{e(rankV2)}				& rank of \stcmd{e(V2)}				\\
\stcmd{e(ll)}				& log-likelihood					& \stcmd{e(ll\_0)}				& log-likelihood, constant-only			\\
\stcmd{e(fpc)}				& finite population correction		&							& \quad model						\\
						& \quad factor					&							&								\\

\stresultsgroup{Macros} \\
\stcmd{e(cmd)}				& \stcmd{probitfe/logitfe}			& \stcmd{e(cmdline)}				& command as typed				\\
\stcmd{e(depvar)}			& name of dependent variable		& \stcmd{e(title)}				& title in estimation output				\\
\stcmd{e(title1)}				& type of included effects			& \stcmd{e(title2)}				& type of correction					\\
\stcmd{e(title3)}				& trimming parameter or \#		& \stcmd{e(chi2type)}			& \stcmd{LR}; type of model chi-squared	\\
						& \quad of multiple partitions		&							& \quad test						\\
\stcmd{e(properties)}			& \stcmd{b V}					& \stcmd{e(id)}					& name of cross-section				\\
\stcmd{e(time)}				& name of time variable			&							& \quad variable					\\

\stresultsgroup{Matrices} \\
\stcmd{e(b)}				& coefficient vector				& \stcmd{e(b2)}					& average partial effects				\\
\stcmd{e(V)}				& variance-covariance matrix		& \stcmd{e(V2)}					& variance-covariance matrix of			\\
						& \quad of coefficient vector		&							& \quad average partial effects			\\

\stresultsgroup{Functions} \\                          
\stcmd{e(sample)}			& marks estimation sample		&							&								\\
\end{stresults}

\section{Bilateral Trade Flows Between Countries}

\subsection{Empirical Example}

To illustrate the use of the bias corrections described in sections 2.4 and 2.5, we present an empirical application to bilateral trade flows between countries using data from Helpman, Melitz and Rubinstein (\citeyear{Helpman01052008}). The data set includes trade flows for 158 countries over the period from 1970 to 1997, as well as country-level data on geography, institutions, and culture (the variables used in the analysis are described below).  We estimate probit and logit models for the probability of positive trade between country pairs in 1986. The data structure is a pseudo-panel where the two dimensions index countries,  {\tt id} as importers  and {\tt jd} as exporters.  There are 157 $\times$ 156 = 24,649 possible country pairs.\footnote{The original data set included 158 countries, but we dropped Congo because it did not export to any country in 1986.}


 For each country pair, the outcome variable {\tt trade}$_{ij}$ is an indicator equal to one if country $i$ imports from country $j$, and equal to zero otherwise. We use $j$ instead of $t$ to emphasize that the second dimension does not index time.  The model specification is based on the gravity equation of \cite{AW2003} with various measures of trade barriers/enhancers as key determinants of international trade flows. We also include the presence of bilateral trade in 1985 to account for possible state dependence in trade decisions. Importer and exporter country fixed effects control for unobserved country heterogeneity such as size, natural resources or trade openness.
The probability that country $i$ imports from country $j$, conditional on the observed variables, $X_{ij}$, the unobserved importer fixed effect, $\alpha_i$, and the unobserved exporter fixed effect, $\gamma_j$, is modeled as
\begin{equation}
\mathbf{Pr}({\tt trade}_{ij}=1 \mid X_{ij}, \alpha_i, \gamma_j) = F(X_{ij}'\beta + \alpha_i + \gamma_j),
\end{equation}
where $F(\cdot)$ is the standard normal cumulative distribution function for the probit model, or the logistic distribution for the logit model. 

The set of explanatory variables, $X_{ij}$, includes:
\begin{enumerate}
\item ${\tt ltrade}_{ij}$: a binary variable equal to one if country $i$ imported from country $j$ in 1985, equal to zero otherwise;
\item ${\tt ldist}_{ij}$: the logarithm of the distance (in km) between country $i$ and country $j$ capitals;
\item ${\tt border}_{ij}$: a binary variable equal to one if country $i$ and country $j$ share a common physical boundary, equal to zero otherwise;
\item ${\tt legal}_{ij}$: a binary variable equal to one if country $i$ and country $j$ share the same legal origin (including civil law, common law, customary law, mixed or pluralistic law, and religious law), equal to zero otherwise;
\item ${\tt language}_{ij}$: a binary variable equal to one if country $i$ and country $j$ share the same official language, equal to zero otherwise;
\item ${\tt colony}_{ij}$: a binary variable equal to one if country $i$ ever colonized country $j$ or vice versa, equal to zero otherwise;
\item ${\tt currency}_{ij}$: a binary variable equal to one if country $i$ and country $j$ use the same currency or if within the country pair money was interchangeable at a 1:1 exchange rate for an extended period of time, equal to zero otherwise;
\item ${\tt fta}_{ij}$: a binary variable equal to one if country $i$ and country $j$ belong to a common regional trade agreement, equal to zero otherwise;
\item ${\tt islands}_{ij}$: a binary variable equal to one if both country $i$ and country $j$ are islands, equal to zero otherwise;
\item ${\tt religion}_{ij}$: the sum of (\% Protestants in country $i$ $\times$ \% Protestants in country $j$) + (\% Catholics in country $i$ $\times$ \% Catholics in country $j$) + (\% Muslims in country $i$ $\times$ \% Muslims in country $j$); and
\item ${\tt landlock}_{ij}$: a binary variable equal to one if both country $i$ and country $j$ have no coastline or direct access to sea, equal to zero otherwise.
\end{enumerate}
The specification of $X_{ij}$ is the same as in Table I of Helpman et al. (\citeyear{Helpman01052008}), except that we include ${\tt ltrade}_{ij}$. Note that despite the inclusion of the lag dependent variable, $X_{ij}$  can be treated as strictly exogenous, because none of the two dimensions of the panel indexes time.

Tables \ref{table:logit} and \ref{table:probit} show the results of the logit model and probit model, respectively. In both tables, column (1) reports uncorrected fixed effect estimates, column (2) reports estimates of the analytical correction setting the trimming parameter equal to zero ({\tt an-0}), and columns (3) to (5) show estimates of the {\tt ss2}, {\tt jj} and {\tt double} jackknife corrections. The {\tt double} correction makes sense because both dimensions of the panel index the same set of countries. Each table shows estimates of index coefficients and APEs. The latter are reported in brackets. We also include standard errors for the index coefficients in column (6), and standard errors for the APEs in columns (6) and (7). In the case of the APEs, the standard errors in column (7) are adjusted by the finite population correction parameter described in Section 3.2, using a population equal to the sample size (24,492). There is only one set of standard errors because the standard errors for the uncorrected estimator are consistent for the corrected estimators (see Fernandez-Val and Weidner (\citeyear{Fernandez-ValandWeidner})). 

\begin{center}
[TABLES \ref{table:logit} AND  \ref{table:probit}  ABOUT  HERE]
\end{center}

We focus on the results for the logit model.  The conclusions from the probit model are analogous, specially  in terms of APEs, which, unlike index coefficients, are comparable  across models. As shown in column (1), the probability that country $i$ imports from country $j$ is higher if country $i$ already imported from country $j$ in the previous year ({\tt ltrade}), if the two countries are closer to each other ({\tt ldist}), if they share the same language ({\tt language}), if they share the same currency ({\tt currency}), if they belong to the same regional free trade agreement ({\tt fta}), if they are not islands ({\tt islands}), or if they share the same religion ({\tt religion}). As in Helpman et al. (\citeyear{Helpman01052008}), the probability that country $i$ imports from country $j$ decreases if both countries have a common land border ({\tt border}), which they attribute to the effect of territorial border conflicts that suppress trade between neighbors. These effects go in the same direction irrespective the type of correction used. However, there are some differences in the  magnitudes of the effects produced by the different estimators.

Comparing across columns, {\tt an-0}, {\tt jk-jj} and {\tt double} produce very similar estimates of index coefficients and APEs, which are all within one standard error of each other.  The split-panel correction estimates of the index coefficients and APEs of {\tt ldist}, {\tt legal}, {\tt currency}, and {\tt fta} in column (3) are two or more standard errors away from the rest of the estimates in the same rows. We show in the next section that  {\tt jk-ss2} is less accurate than {\tt an-0} and {\tt double} through a Monte Carlo simulation calibrated to this application. Relative to the uncorrected estimates in column (1),  the corrected estimates of the index coefficient of ${\tt ltrade}$ are more than one standard error lower.  We attribute the similarity in the rest of index coefficients and APES between uncorrected and bias corrected estimates partly to the large sample size (except for {\tt jk-ss2}). Thus, we find more significant differences  in the next section when we consider subpanels with less than 157 countries.

\subsection{Calibrated Monte Carlo Simulations}

To evaluate the performance of the bias corrections, we  conduct a Monte Carlo simulation that mimics the empirical example described above. We focus on the logit model, leaving the probit model for the online supplementary material. All the parameters are calibrated to the data used in the previous section, and their values are set to the uncorrected fixed effect estimates from column (1) in Table \ref{table:logit}. To speed up computation, we consider only two explanatory variables in $X_{ij}$: the presence of trade in the previous year ({\tt ltrade}) and the log distance between country pairs ({\tt ldist}).

For all possible country pairs we first construct the index
\begin{equation*}
{\tt index}_{ij} = \widehat{\beta}_1{\tt ltrade}_{ij} + \widehat{\beta}_2{\tt ldist}_{ij} + \widehat{\alpha}_i + \widehat{\gamma}_j ,
\end{equation*}
where $\widehat{\beta}_1=2.838$, $\widehat{\beta}_2=-0.839$, and $\widehat{\alpha}_i$ and $\widehat{\gamma}_j$ are the uncorrected estimates of the importer and exporter fixed effects (not reported in Table \ref{table:logit}). Next, we generate a new trade indicator for each country  pair as
\begin{equation*}
{\tt trade}_{ij}^* = 1 \cdot \Bigg\{ {\tt index}_{ij} > \underbrace{ {\tt ln} \left( \frac{1} { {\tt runiform}(1,1)} - 1 \right)}_{(*)} \Bigg\} ,
\end{equation*}
where {\tt ln} denotes the natural logarithm and {\tt runiform}(1,1) generates a random number from the uniform distribution in $(0,1)$, such that $(*)$ corresponds to a random draw from the standard logistic distribution. 


We use the generated trade indicators to estimate the equation
\begin{equation*}
\mathbf{Pr}({\tt trade}_{ij}^*=1 \mid {\tt ltrade}_{ij}, {\tt ldist}_{ij}, \alpha_i, \gamma_j) = F(\beta_1^*{\tt ltrade}_{ij} + \beta_2^*{\tt ldist}_{ij}  + \alpha^*_i + \gamma^*_j) ,
\end{equation*}
where $F(\cdot)$ is the logistic distribution, and {\tt ltrade} and {\tt dist} are the variables from the original data set. We repeat this procedure in 500 simulations for five different sample sizes: $N=25$, $N=50$, $N=75$, $N=100$ and $N=157$ (full sample). For each sample size and simulation, we draw a random sample of $N$ countries both as importers and exporters without replacement, so that the number of observations is $N\times (N-1)$.

Table \ref{table:mc} reports the result for the uncorrected estimator (FE), analytical correction setting the trimming parameter equal to zero (AN-0), jackknife correction {\tt ss2} (JK-SS2), and jackknife correction {\tt double} (Double). We analyze the performance of these estimators in terms of bias and inference accuracy of their asymptotic distribution for both index coefficients and APEs. In particular, we compute the biases (Bias), standard deviations (Std. Dev.),  and root mean squared errors (RMSE) of the estimators, together with the ratio of average standard errors to the simulation standard deviations (SE/SD), and the empirical coverages of confidence intervals with 95\% nominal level (p; .95). The variance of the APEs is adjusted by the {\tt \dunderbar{pop}ulation({\it integer})} option, with the population being equal to the original sample size (24,492 observations). All the results are reported in percentage of the true parameter value.

\begin{center}
[TABLE \ref{table:mc} ABOUT HERE]
\end{center}

For the uncorrected estimators in Panel A, we observe in column (1) that there is significant bias in the index coefficients. This bias decreases with the sample size, but it is still larger than the standard deviation for the coefficient of {\tt ltrade} in the full sample.  Moreover, column (5) shows that confidence intervals constructed around the uncorrected estimates suffer from severe undercoverage for all the sample sizes. As in Fernandez-Val and Weidner (\citeyear{Fernandez-ValandWeidner}), we find very little bias in the APEs, despite the large bias in the index coefficients.  In Panel B, we see that the analytical correction reduces substantially the bias in the index coefficients producing confidence intervals with coverage close to their nominal level for every sample size.  This correction reduces standard deviation resulting in a reduction of more than $50\%$ in rmse for several sample sizes. The jackknife corrections also reduce bias and generally improve coverage, but increase dispersion in small samples and require of larger sample sizes than the analytical corrections to improve rmse over the uncorrected estimator.   The jackknife correction {\tt double} performs very similarly to the analytical correction except for the smallest sample size. The jackknife correction {\tt ss2} of the index coefficient of {\tt ldist} has higher rmse than the uncorrected estimator even for the full sample size. 
Overall, the standard errors provide a good approximation to the standard deviations of all the estimators of both the index coefficients and APEs.


To sum up, Table \ref{table:mc} shows that the analytical correction substantially reduces the bias of the uncorrected estimator, producing more accurate point and interval estimators for all the sample sizes considered. The jackknife correction {\tt double} performs similarly to the analytical correction, except for the smallest sample size $N=25$.  The split panel correction {\tt ss2} reduces bias, but at the cost of increasing dispersion for most sample sizes. In this application {\tt ss2} is dominated by the other corrections uniformly across all the sample sizes in terms of rmse. These results are consistent with the empirical evidence in Table \ref{table:logit}, where the uncorrected estimates of the index coefficient of {\tt ltrade} were more than one standard error below the corrected estimates, the estimates of the APEs where very similar for the uncorrected and corrected estimators except for {\tt ss2}, and the jaccknife correction {\tt ss2} produced estimates for {\tt ldist} at odds with the other estimators.

\section{Concluding remarks}

The commands  {\tt probitfe} and {\tt logitfe} implement the analytical and jackknife bias corrections of Fernandez-Val and Weidner (\citeyear{Fernandez-ValandWeidner}) for logit and probit models with two-way fixed effects. The commands compute estimators of both index coefficients and APEs, which are often the parameters of interest in these models. We also provide functionality for models with one-way fixed effects, offering an alternative to the commands \rref{clogit} and \rref{xtlogit} that do not produce corrected estimates of APEs. Logit and probit models are commonly used in empirical work, making the new commands a valuable addition to the applied econometricianÕs toolkit. Similar corrections can be implemented for other nonlinear panel models such as tobit models for censored outcome variables. We leave this extension to future research.

\section{Acknowledgments}
 
We thank an anonymous reviewer for helpful comments and suggestions. Mario Cruz-Gonzalez gratefully acknowledges support from the National Science Foundation grant No. SES-1060889.
Iv\'an Fern\'andez-Val gratefully acknowledges support from the National Science Foundation grant No. SES-1060889. 
Martin Weidner gratefully acknowledges support from the Economic and Social Research Council through the ESRC Centre for Microdata Methods and Practice grant RES-589-28-0001,
and also from the European Research Council (ERC) grant ERC-2014-CoG-646917-ROMIA.

\section{Appendix}

\subsection{Expressions of the asymptotic  bias and variance}
Fernandez-Val and Weidner  (\citeyear{Fernandez-ValandWeidner}) show that the asymptotic bias and variance for $\beta$ can be expressed as
$$
 B^{\beta} =  W^{-1}  B, \ \  D^{\beta} =  W^{-1}  D, \  \  V^{\beta} =  W^{-1}, 
$$
where
\begin{eqnarray*}
 B &=&    \EE \left[ -
         \frac {1} {2N}  \sum_{i=1}^{N}
            \frac{  \sum_{t=1}^T \left\{ H_{it} \partial^2 F_{it} \tilde{X}_{it} +  2 \sum_{\tau=t+1}^T
         H_{it}
               (Y_{it} - F_{it} ) \omega_{i\tau}  \tilde{X}_{i\tau}    \right\} } {  \sum_{t=1}^T  \omega_{it}  } \right],\\
 D &=&  \EE \left[ -
         \frac {1} {2T}  \sum_{t=1}^{T}
            \frac{  \sum_{i=1}^N  H_{it} \partial^2 F_{it} \tilde{X}_{it}} {  \sum_{i=1}^N  \omega_{it} }\right] ,   \\          
 W &=&     \EE \left[
         \frac {1} {NT}  \sum_{i=1}^{N} \sum_{t=1}^{T}
           \omega_{it} \tilde{X}_{it} \tilde{X}_{it}'  \right],
\end{eqnarray*}
$\EE := \plim_{N,T \rightarrow \infty}$, $\omega_{it} = H_{it} \partial F_{it}$, $H_{it} = \partial F_{it} / [F_{it}(1-F_{it})],  $   $\partial^{j}  G_{it} := \partial^{j} G(Z)|_{Z = X_{it}'\beta^0 + \alpha_{i}^0 + \gamma_t^0}$ for any function $G$ and $j = 0,1,2$,  and $\tilde X_{it}$ is the residual of the population projection of $X_{it}$ on the space spanned by $\alpha_i$  and $\gamma_t$ under a metric weighted by $\omega_{it}$.  

The expressions of the asymptotic bias terms for the APEs are different depending on whether the APEs are obtained from uncorrected or bias corrected estimators of $\beta$.  The commands \rref{probitfe} and \rref{logitfe} implement the corrections on APEs obtained from bias corrected estimators of the parameters, that is, $\widetilde \delta$ is obtained using
 $\widetilde \beta$ equal to the bias corrected estimator $ \widetilde \beta^A$ defined below.
The expressions for the leading bias terms of $\widetilde \delta$ then read
\begin{align*}
     B^{\delta} &=  \EE \left[  \frac {1} {2N}  \sum_{i=1}^{N} 
            \frac{  \sum_{t=1}^T \{  2 \sum_{\tau=t+1}^T
          H_{it}(Y_{it} - F_{it})  \omega_{i\tau} \tilde \Psi_{i\tau}
                   + \partial_{\alpha_i^2} \Delta_{it}  - \Psi_{it} H_{it} \partial^2 F_{it} \}}
        {  \sum_{t=1}^T \omega_{it} }\right],
       \\
 D^{\delta} &= \EE \left[
         \frac {1} {2T}  \sum_{t=1}^{T}
            \frac{  \sum_{i=1}^N
         \{   \partial_{\alpha_i^2} \Delta_{it}  - \Psi_{it} H_{it} \partial^2 F_{it}  \}}
        {  \sum_{i=1}^N  \omega_{it} } \right],
\end{align*}
where $\Psi_{it}$ and $\tilde \Psi_{it}$ are the fitted value and residual of the population regression of
$ - \partial_{\pi} \Delta_{it} /
                        \omega_{it}$ 
                        on the space spanned by $\alpha_i$ and $\gamma_t$ under the
metric given by $\omega_{it}$.  If all the components of $X_{it}$  are strictly exogenous,
the first term of $  B^{\delta}$ is zero. The asymptotic variance of the estimators of $\delta$ is
$$
          {V}^{\delta} = \EE \left\{
       \frac {r_{NT}^2} { N^2 T^2}
       \sum_{i=1}^N \left[ \sum_{t, \tau = 1}^T  \widetilde \Delta_{it} \widetilde \Delta_{i \tau}'  + \sum_{j \neq i}\sum_{t = 1}^T  \widetilde \Delta_{it} \widetilde \Delta_{j t} ' +  \sum_{t=1}^T \Gamma_{it} \Gamma_{it}'\right] \right\},
$$
where $r_{NT} = \sqrt{NT/(N+T-1)}$, $ \widetilde \Delta_{it} = \Delta_{it} - \delta^0$,  $\Gamma_{it}=   {(D_{\beta} \Delta)}'  W_\infty^{-1} H_{it} (Y_{it} - F_{it}) \tilde X_{it}
      -   \Psi_{it} 
         H_{it} (Y_{it} - F_{it}) $, and
         $$
                 {D_{\beta} \Delta} = \EE \left[
       \frac 1 {NT} \sum_{i=1}^N \sum_{t=1}^T
          \partial_{\alpha_i} \Delta_{it} \tilde X_{it}    \right].
         $$ 

\subsection{Analytical Correction}\label{sec:abc}

The analytical corrections are implemented using plug-in estimators of the bias terms that replace expectations by sample averages and  true parameter values by fixed effect estimators. Thus,  for any function of the data, unobserved effects and parameters $g_{it}(\beta,\alpha_i, \gamma_t)$, let $\widehat g_{it} = g_{it}(\widehat \beta, \widehat \alpha_i, \widehat \gamma_t)$ denote the fixed effect estimator of $g_{it} = g_{it}(\beta^0,\alpha_i^0, \gamma_t^0)$, e.g., $\widehat{F}_{it} = F(X_{it}'\widehat{\beta} + \widehat \alpha_i + \widehat \gamma_t )$ denotes the fixed effect estimator of $F_{it} = F(X_{it}'\beta^0 + \alpha_i^0 +  \gamma_t ^0).$ The commands {\tt probitfe} and {\tt logitfe} with the {\tt \underbar{an}alytical} option compute the correction for $\beta$
$$
\widetilde \beta^A = \widehat \beta - \widehat W^{-1} \widehat B/T - \widehat W^{-1}  \widehat D/N,
$$
where
\begin{eqnarray*}
\widehat B &=&   \textstyle   - 
         \frac {1} {2N}  \sum_{i=1}^{N}
            \frac{  \sum_{t=1}^T  \widehat H_{it} \partial^2 \widehat F_{it} \widehat{\tilde{X}}_{it} + 2 \sum_{j=1}^L  [T/(T-j)] \sum_{t=j+1}^{T}
       \widehat H_{i,t-j}
               (Y_{i,t-j} - \widehat F_{i,t-j} ) \widehat \omega_{it}  \widehat{\tilde{X}}_{it} } {  \sum_{t=1}^T \widehat \omega_{it} } , \\
\widehat D &=&    -
         \frac {1} {2T}  \sum_{t=1}^{T}
            \frac{  \sum_{i=1}^N  \widehat H_{it} \partial^2 \widehat F_{it} \widehat{\tilde{X}}_{it} } {  \sum_{i=1}^N \widehat \omega_{it} } ,   \\          
\widehat W &=&   
         \frac {1} {NT}  \sum_{i=1}^{N} \sum_{t=1}^{T}
            \widehat \omega_{it} \widehat{\tilde{X}}_{it} \widehat{\tilde{X}}_{it}',
\end{eqnarray*}
$\widehat \omega_{it} = \widehat H_{it} \partial \widehat F_{it}$, $\widehat{\tilde X}_{it}$ is the residual of the least squares  projection of $X_{it}$ on the space spanned by the incidental parameters under a metric weighted by $\widehat \omega_{it}$, and $L$ is a trimming parameter for estimation of spectral expectations  such that $L \to \infty$ and $L/T \to 0$. The factor $T/(T-j)$ is a degrees of freedom adjustment that rescales the time series averages $T^{-1} \sum_{t=j+1}^T$ by the number of observations instead of by $T$.

Similarly, the analytical correction for $\delta$ is computed as
$$
\widetilde \delta^A = \widetilde \delta -  \widehat B^{\delta}/T -   \widehat D^{\delta}/N,
$$
where
\begin{align*}
\widetilde \delta &= \Delta(\widetilde \beta^A, \widetilde \alpha^A, \widetilde \gamma^A),\\
(\widetilde \alpha^A, \widetilde \gamma^A) &\in \argmax_{(\alpha,\gamma) \in \mathbb{R}^{N +T}} \sum_{i,t}  \ell_{it}(\widetilde \beta^A, \alpha_i, \gamma_t), \\
    \widehat B^{\delta} &=  \textstyle \frac {1} {2N}  \sum_{i=1}^{N} 
            \frac{   2 \sum_{j=1}^L  [T/(T-j)] \sum_{t=j+1}^{T}
          \widehat H_{i,t-j}(Y_{i,t-j} - \widehat F_{i,t-j}) \widehat \omega_{it} \widehat{\tilde \Psi}_{it}
                + \sum_{t=1}^T \{   \partial_{\alpha_i^2} \widehat \Delta_{it}  - \widehat \Psi_{it} \widehat H_{it} \partial^2 \widehat F_{it}  \}}
        {  \sum_{t=1}^T \widehat \omega_{it} },
       \\
\widehat D^{\delta} &=  
         \frac {1} {2T}  \sum_{t=1}^{T}
            \frac{  \sum_{i=1}^N
         \{   \partial_{\alpha_i^2} \widehat \Delta_{it}  - \widehat \Psi_{it} \widehat H_{it} \partial^2 \widehat F_{it}  \}}
        {  \sum_{i=1}^N  \widehat \omega_{it} }.
\end{align*}

\subsection{Standard Errors}\label{sec:se}

The standard errors for all the estimators (uncorrected or corrected) of the $k^{\rm th}$ component of $\beta$ are computed as
$$
\sqrt{\widehat W_{kk}^{-1} / (NT)},  \ \ k = \{1, ..., \dim \beta\},
$$
where $\widehat W_{kk}^{-1} $ is the $(k,k)$-element of the  matrix $\widehat W^{-1}$ defined above, which is based on the uncorrected fixed effect estimator $\widehat \beta$.  The standard errors for all the estimators of the APEs are computed as
 \begin{equation*}
       \frac {1} {NT} \left\{
       \sum_{i=1}^N  \left[  a_{NT} \sum_{t,\tau = 1}^T \widehat{\tilde \Delta}_{it} \widehat{\tilde \Delta}_{i\tau}' +  a_{NT} \sum_{t = 1}^T \sum_{j \neq i}  \widehat{\tilde \Delta}_{it} \widehat{\tilde \Delta}_{jt}' + \sum_{t=1}^{T} \widehat \Gamma_{it} \widehat \Gamma_{it}'  \right]\right\}^{1/2},
       \end{equation*}
where $ \widehat{\tilde \Delta}_{it} = \widehat \Delta_{it} - \widetilde \delta$, $\widehat \Gamma_{it}=  (D_{\beta} \widehat{\Delta})' \widehat W^{-1} \widehat H_{it} (Y_{it} - \widehat F_{it}) \widehat{\tilde X}_{it}
      -   \widehat \Psi_{it} 
         \widehat H_{it} (Y_{it} - \widehat F_{it}) $, and
         $$
                D_{\beta} \widehat{\Delta} = 
       \frac 1 {NT} \sum_{i=1}^N \sum_{t=1}^T
          \partial_{\alpha_i} \widehat \Delta_{it} \widehat{\tilde X}_{it},
         $$ 
The factor $a_{NT}$ is a finite population correction term, 
$$a_{NT} = (N_0T_0 - NT)/(N_0 T_0 - 1),$$ 
where $N_0$ and $T_0$ are the population sizes of the 2 dimensions of the panel.  For example,  $a_{NT} = 1$ if at least one of the dimension has infinite  size in the population, and $a_{NT} = 0$ if we observe the entire population. The correction only affects the first two terms of the variance because they come from using a sample mean to estimate a population mean, whereas the third term is due to parameter estimation.

\subsection{One-Way Fixed Effect Models}

In models that include only individual effects, all the expressions of the asymptotic bias and variance are the same as for the two-way fixed effect models except for
$$
D^{\beta} =  0,  \ \ D^{\delta} = 0, \ \  \partial^{j}  G_{it} := \partial^{j} G(Z)|_{Z = X_{it}'\beta^0 + \alpha_{i}^0},
$$
and $\tilde X_{it}$ is the residual of the population projection of $X_{it}$ on the space spanned by $\alpha_i$  under a metric weighted by $\omega_{it}$.  Symmetrically, in models that include only time  effects, all the expressions of the asymptotic bias and variance are the same as for the two-way fixed effect models except for
$$
B^{\beta} =  0, \ \  B^{\delta} = 0, \ \  \partial^{j}  G_{it} := \partial^{j} G(Z)|_{Z = X_{it}'\beta^0  + \gamma_t^0},
$$
and $\tilde X_{it}$ is the residual of the population projection of $X_{it}$ on the space spanned by $\gamma_t$  under a metric weighted by $\omega_{it}$. 

We do not provide explicit expressions for the analytical bias corrections and standard errors because they are analogous to the expressions given in  Sections \ref{sec:abc} and \ref{sec:se}. For the jackknife, in models that include only individual effects: 
\begin{itemize}
\item
The corrections {\tt ss1}, {\tt ss2} and {\tt sj} implement the SPJ of Dhaene and Jochmans that applies SPJ to the individual dimension, that is
\begin{equation*}
\widetilde{\beta}^{\rm ss1} = \widetilde{\beta}^{\rm ss2} =  \widetilde{\beta}^{\rm sj}  = 2  \widehat \beta - \widetilde{\beta}_{N,T/2}.
\end{equation*}

\item

The corrections {\tt js}, {\tt jj}, and {\tt double} implement the jackknife correction of Hahn and Newey that applies PJ to the individual dimension, that is
\begin{equation*}
\widetilde{\beta}^{\rm js} = \widetilde{\beta}^{\rm jj}  = \widetilde{\beta}^{\rm double}  = N \widehat{\beta} - (N-1)\widetilde{\beta}_{N-1,T}.
\end{equation*}
\end{itemize}
Similarly, in models that include only time effects: 
\begin{itemize}
\item
The corrections {\tt ss1}, {\tt ss2},  and  {\tt js} implement the SPJ of Dhaene and Jochmans that applies SPJ to the time dimension, that is
\begin{equation*}
\widetilde{\beta}^{\rm ss1} = \widetilde{\beta}^{\rm ss2}  = \widetilde{\beta}^{\rm js} =  2  \widehat \beta - \widetilde{\beta}_{N/2,T}.
\end{equation*}

\item

The corrections {\tt sj}, {\tt jj} and {\tt double} implement the jackknife correction of Hahn and Newey that applies PJ to the time dimension, that is
\begin{equation*}
\widetilde{\beta}^{\rm sj} = \widetilde{\beta}^{\rm jj}  = \widetilde{\beta}^{\rm double}  = T \widehat{\beta} - (T-1)\widetilde{\beta}_{N,T-1}.
\end{equation*}
\end{itemize}

\bibliographystyle{sj}
\bibliography{references}

\begin{aboutauthors}
    Mario Cruz-Gonzalez is a PhD candidate in the Department of Economics at Boston University. His research interests include topics in labor economics, development economics and econometrics.

  Iv\'an Fern\'andez-Val is an Associate Professor of Economics at Boston University. His research interests focus on topics in theoretical and applied econometrics, including panel data analysis, and distributional and quantile methods.

  Martin Weidner is a Lecturer in Economics at University College London,
  and a member of the Centre for Microdata Methods and Practice (CeMMAP)
  at the Institute for Fiscal Studies in London.
  He is working on Theoretical and Applied Econometrics, with a special focus on Panel Data, Factor Models, Demand Estimation, and Social Networks.
  
\end{aboutauthors}

\begin{table}[!hbt]\footnotesize
\centering
\begin{threeparttable}
\def\sym#1{\ifmmode^{#1}\else\(^{#1}\)\fi}
\caption[center,footnotesize]{{\bf Fixed Effect Logit Model}}\label{table:logit}
\begin{tabular}{l*{7}{c}}
\hline
\toprule
                    &\multicolumn{1}{c}{(1)}&\multicolumn{1}{c}{(2)}&\multicolumn{1}{c}{(3)}&\multicolumn{1}{c}{(4)}&\multicolumn{1}{c}{(5)}&\multicolumn{1}{c}{(6)}&\multicolumn{1}{c}{(7)}\\
                    &\multicolumn{1}{c}{FE}&\multicolumn{1}{c}{AN-0}&\multicolumn{1}{c}{JK-SS2}&\multicolumn{1}{c}{JK-JJ}&\multicolumn{1}{c}{Double}&\multicolumn{2}{c}{Std. Error}\\
\midrule
{\tt ltrade}              &       2.838&       2.741&        2.786&       2.743&       2.745 & (0.058)\\
 &       [0.325]&       [0.323]&          [0.349]&        [0.325]&       [0.326] & (0.014) & (0.008)\\                 
[1em]
{\tt ldist}               &      -0.839&      -0.819&      -0.742&       -0.812&      -0.812 & (0.044)\\
&      [-0.055]&      [-0.055]&      [-0.049]&        [-0.055]&      [-0.055] & (0.004) & (0.003)\\
[1em]
{\tt border}              &      -0.571&      -0.557&        -0.493&       -0.564&      -0.573 & (0.195)\\
 &      [-0.037] &     [ -0.037] &        [-0.036]  &          [-0.037] &      [-0.038] & (0.012) & (0.012)\\                   
[1em]
{\tt legal}               &       0.115         &       0.113            &       0.017        &       0.112         &       0.112       & (0.062)  \\
  &       [0.008]         &       [0.008]          &       [0.003]               &       [0.008]         &       [0.008]       & (0.004)  & (0.004)\\                  
[1em]
{\tt language}            &       0.368&       0.358&      0.385&      0.354&       0.352 & (0.080)\\
 &       [0.025]&       [0.025]&         [0.026]&         [0.024]&       [0.024] & (0.005) & (0.005)\\                   
[1em]
{\tt colony}              &       0.492         &       0.435             &      -0.023         &       0.344         &       0.129    & (0.633)     \\
 &       [0.034]         &       [0.030]            &        [0.002]         &       [0.021]         &       [0.004]  & (0.045)     & (0.045)  \\                   
[1em]
{\tt currency}            &       0.984&       0.961&       2.464&       1.009&       1.079 & (0.252)\\
  &       [0.070]&       [0.070]&       [0.164]&       [0.071]&       [0.073] & (0.020) & (0.019)\\                   
[1em]
{\tt fta}                 &       2.244&       2.171&         3.347&     1.827&       1.571 & (0.657)\\
  &       [0.178]&       [0.177]&     [0.285]&         [0.142] &       [0.118] & (0.062) & (0.061)\\                   
[1em]
{\tt islands}            &       0.406&       0.395 &         0.393  &       0.396  &       0.396 & (0.156)\\
&       [0.027] &       [0.027]  &         [0.028]  &          [0.027]  &       [0.027] & (0.011) & (0.011)\\                    
[1em]
{\tt religion}            &       0.244&       0.239          &       0.238    &       0.240         &       0.245 & (0.123)\\
 &       [0.016]  &       [0.016]         &       [0.017]      &       [0.016]         &       [0.017]  & (0.008) & (0.008)\\                   
[1em]
{\tt landlock}           &       0.143         &       0.139             &       0.153     &       0.156         &       0.170        & (0.221) \\
 &       [0.010]         &       [0.010]        &       [0.014]            &       [0.010]         &       [0.011]        & (0.015)  & (0.015)\\
 \midrule
Obs.        &       24492         &       24492             &       24492             &       24492         &       24492         \\
\bottomrule
\hline
\end{tabular}
\begin{tablenotes}
\footnotesize
\item[] {\it Notes}: Average Partial Effects in brackets. FE denotes uncorrected fixed effect estimator; AN-0 denotes analytical correction with 0 lags; JK-SS2 denotes split jackknife in both dimensions;  JK-JJ denotes delete-one jackknife in both dimensions; Double denotes delete-one jackknife for observations with the same index in the cross-section and the time-series. For the Average Partial Effects, the standard errors reported in Column (7) are adjusted by the finite population correction parameter, using a population equal to the number of observations (24,492).
\end{tablenotes}
\end{threeparttable}
\end{table}

\begin{table}[!hbt]\footnotesize
\centering
\begin{threeparttable}
\def\sym#1{\ifmmode^{#1}\else\(^{#1}\)\fi}
\caption[center,footnotesize]{{\bf Fixed Effect Probit Model}}\label{table:probit}
\begin{tabular}{l*{7}{c}}
\hline
\toprule
                    &\multicolumn{1}{c}{(1)}&\multicolumn{1}{c}{(2)}&\multicolumn{1}{c}{(3)}&\multicolumn{1}{c}{(4)}&\multicolumn{1}{c}{(5)}&\multicolumn{1}{c}{(6)}&\multicolumn{1}{c}{(7)}\\
                    &\multicolumn{1}{c}{FE}&\multicolumn{1}{c}{AN-0}&\multicolumn{1}{c}{JK-SS2}&\multicolumn{1}{c}{JK-JJ}&\multicolumn{1}{c}{Double}&\multicolumn{2}{c}{Std. Error}\\
\midrule
{\tt ltrade}              &       1.631&       1.586&           1.625&            1.587&       1.588 & (0.031) &\\
&       [0.343]&       [0.345]&           [0.371]&           [0.346]&       [0.347] & (0.014) & (0.009)\\                   
[1em]
{\tt ldist}               &      -0.438&      -0.426&          -0.377&         -0.423&      -0.422 & (0.023)& \\
&      [-0.054]&      [-0.054]&          [-0.046]&         [-0.054]&      [-0.054] & (0.004)& (0.003)\\                   
[1em]
{\tt border}              &      -0.273  &      -0.265  &         -0.208         &          -0.268  &      -0.273  & (0.107)&\\
&     [ -0.033]&     [ -0.033]&           [-0.029]&         [-0.033]&      [-0.034]  & (0.013)&(0.012)\\                    
[1em]
{\tt legal}               &       0.059         &       0.057                 &       0.011             &       0.056         &       0.056       & (0.033)  &\\
&       [0.007]&       [0.007]&            [0.003]&       [0.007]&       [0.007]     & (0.004)  & (0.004)  \\                    
[1em]
{\tt language}            &       0.203&       0.198&         0.215&          0.196&       0.196 & (0.042)&\\
&       [0.025]&       [0.025]&           [0.027]&          [0.025]&       [0.025] & (0.005)&(0.005)\\                    
[1em]
{\tt colony}              &       0.287         &       0.253               &       0.005 &       0.207         &       0.099       & (0.356)  \\
&       [0.037]&       [0.033]&           [0.006]&         [0.025]&       [0.008]    & (0.047)  & (0.047)   \\                    
[1em]
{\tt currency}            &       0.529&       0.515&            1.340&            0.537&       0.568 & (0.139)&\\
&       [0.069]&       [0.070]&            [0.166]&          [0.070]&       [0.072] & (0.020)&(0.019)\\                    
[1em]
{\tt fta}                 &       1.235&       1.192&           1.807&            1.067 &       0.991 & (0.340)&\\
&       [0.180]&       [0.178]&             [0.281]&          [0.155]&       [0.143] & (0.057) &(0.057)\\                    
[1em]
{\tt islands}            &       0.194  &       0.187  &          0.203  &           0.188  &       0.188 & (0.084) &\\
&       [0.024]&       [0.024]&            [0.026]&          [0.024]&       [0.024] & (0.011)&(0.011) \\                   
[1em]
{\tt religion}            &       0.134  &       0.132  &            0.133  &          0.133  &       0.135 & (0.066)& \\
&       [0.017]&       [0.017]&              [0.018]&       [0.017]&       [0.017] & (0.008)  &(0.008)\\                   
[1em]
{\tt landlock}           &       0.041         &       0.041                 &           0.033         &       0.044         &       0.049      & (0.119)&   \\
&       [0.005]&       [0.005]&             [0.008]&        [0.005]&       [0.006]  & (0.015)    &(0.015)   \\                    
\midrule
Obs.        &       24492         &       24492                 &              24492         &       24492         &       24492         \\
\bottomrule
\hline
\end{tabular}
\begin{tablenotes}
\footnotesize
\item[] {\it Notes}: Average Partial Effects in brackets. FE denotes uncorrected fixed effect estimator; AN-0 denotes analytical correction with 0 lags; JK-SS2 denotes split jackknife in both dimensions; JK-JJ denotes delete-one jackknife in both dimensions; Double denotes delete-one jackknife for observations with the same index in the cross-section and the time-series. For the Average Partial Effects, the standard errors reported in Column (7) are adjusted by the finite population correction parameter, using a population equal to the number of observations (24,492).
\end{tablenotes}
\end{threeparttable}
\end{table}

\begin{landscape}
\begin{table}[!hbt]\footnotesize
\centering
\begin{threeparttable}
\def\sym#1{\ifmmode^{#1}\else\(^{#1}\)\fi}
\caption[center,footnotesize]{{\bf Calibrated Monte Carlo Simulations for the Logit Model}}\label{table:mc}
\begin{tabular}{ll*{11}{c}}
\hline
\toprule
&&\multicolumn{5}{c}{Index Coefficients}& & \multicolumn{5}{c}{Average Partial Effects}\\
\cmidrule{3-7} \cmidrule{9-13}
&&\multicolumn{1}{c}{(1)}&\multicolumn{1}{c}{(2)}&\multicolumn{1}{c}{(3)}&\multicolumn{1}{c}{(4)}&\multicolumn{1}{c}{(5)}&&\multicolumn{1}{c}{(6)}&\multicolumn{1}{c}{(7)}&\multicolumn{1}{c}{(8)}&\multicolumn{1}{c}{(9)}&\multicolumn{1}{c}{(10)}\\
&&\multicolumn{1}{c}{Bias}&\multicolumn{1}{c}{Std. Dev.}&\multicolumn{1}{c}{RMSE}&\multicolumn{1}{c}{SE/SD}&\multicolumn{1}{c}{p; 95}&&\multicolumn{1}{c}{Bias}&\multicolumn{1}{c}{Std. Dev.}&\multicolumn{1}{c}{RMSE}&\multicolumn{1}{c}{SE/SD}&\multicolumn{1}{c}{p; .95}\\
\midrule
\multicolumn{13}{l}{\bf A. FE}\\
\multirow{2}{*}{N=25}	& {\tt ltrade}	&	33.640	&	24.422	&	41.556	&	0.806	&	0.654	&&	0.662	&	20.430	&	20.421	&	0.925	&	0.928	\\
					& {\tt ldist}		&	27.470	&	46.597	&	54.051	&	0.857	&	0.898	&&	-0.211	&	37.521	&	37.484	&	1.094	&	0.958	\\
[1em]																						
\multirow{2}{*}{N=50}	& {\tt ltrade}	&	12.768	&	8.003	&	15.065	&	0.943	&	0.606	&&	-0.727	&	10.776	&	10.790	&	0.938	&	0.924	\\
					& {\tt ldist}		&	10.629	&	17.095	&	20.115	&	0.949	&	0.902	&&	-1.847	&	17.273	&	17.354	&	1.148	&	0.960	\\
[1em]																						
\multirow{2}{*}{N=75}	&{\tt ltrade}	&	7.681	&	5.132	&	9.235	&	0.914	&	0.614	&&	-1.205	&	7.130	&	7.224	&	0.984	&	0.944	\\
					&{\tt ldist}		&	6.421	&	11.101	&	12.815	&	0.924	&	0.896	&&	-1.766	&	11.869	&	11.988	&	1.121	&	0.962	\\
[1em]																						
\multirow{2}{*}{N=100}	&{\tt ltrade}	&	5.615	&	3.627	&	6.683	&	0.939	&	0.614	&&	-1.184	&	5.058	&	5.189	&	1.042	&	0.946	\\
					&{\tt ldist}		&	4.308	&	8.212	&	9.266	&	0.915	&	0.910	&&	-2.003	&	8.771	&	8.988	&	1.132	&	0.962	\\
[1em]																						
\multirow{2}{*}{N=157}	&{\tt ltrade}	&	3.534	&	2.174	&	4.148	&	0.965	&	0.592	&&	-1.279	&	2.611	&	2.905	&	0.984	&	0.920	\\
					&{\tt ldist}		&	2.694	&	4.555	&	5.288	&	1.025	&	0.920	&&	-1.975	&	4.283	&	4.712	&	1.329	&	0.976	\\
\midrule
\multicolumn{13}{l}{\bf B. AN-0}\\
\multirow{2}{*}{N=25}	& {\tt ltrade}	&	-1.726	&	14.631	&	14.716	&	1.334	&	0.986	&&	-2.178	&	19.547	&	19.643	&	0.965	&	0.922	\\
					& {\tt ldist}		&	-1.672	&	32.615	&	32.621	&	1.219	&	0.986	&&	2.686	&	37.116	&	37.165	&	1.119	&	0.961	\\
[1em]																						
\multirow{2}{*}{N=50}	& {\tt ltrade}	&	0.142	&	6.675	&	6.669	&	1.131	&	0.974	&&	-2.352	&	10.841	&	11.082	&	0.931	&	0.905	\\
					& {\tt ldist}		&	0.570	&	15.271	&	15.266	&	1.062	&	0.966	&&	-1.276	&	17.117	&	17.147	&	1.158	&	0.959	\\
[1em]																						
\multirow{2}{*}{N=75}	&{\tt ltrade}	&	-0.151	&	4.592	&	4.590	&	1.022	&	0.964	&&	-2.358	&	7.174	&	7.544	&	0.977	&	0.929	\\
					&{\tt ldist}		&	0.353	&	10.355	&	10.350	&	0.991	&	0.956	&&	-1.392	&	11.783	&	11.853	&	1.129	&	0.961	\\
[1em]																						
\multirow{2}{*}{N=100}	&{\tt ltrade}	&	-0.090	&	3.331	&	3.329	&	1.022	&	0.948	&&	-2.081	&	5.065	&	5.471	&	1.040	&	0.930	\\
					&{\tt ldist}		&	-0.029	&	7.804	&	7.796	&	0.962	&	0.934	&&	-1.680	&	8.729	&	8.881	&	1.137	&	0.974	\\
[1em]																						
\multirow{2}{*}{N=157}	&{\tt ltrade}	&	-0.020	&	2.069	&	2.067	&	1.014	&	0.972	&&	-1.871	&	2.602	&	3.203	&	0.987	&	0.864	\\
					&{\tt ldist}		&	0.059	&	4.416	&	4.412	&	1.058	&	0.960	&&	-1.732	&	4.281	&	4.614	&	1.329	&	0.982	\\
\midrule
\multicolumn{13}{r}{\it Continued...}
\end{tabular}
\end{threeparttable}
\end{table}
\end{landscape}

\begin{landscape}
\begin{table}[!hbt]\footnotesize
\centering
\begin{threeparttable}
\def\sym#1{\ifmmode^{#1}\else\(^{#1}\)\fi}
\begin{tabular}{ll*{11}{c}}
\hline
\toprule
&&\multicolumn{5}{c}{Index Coefficients}& & \multicolumn{5}{c}{Average Partial Effects}\\
\cmidrule{3-7} \cmidrule{9-13}
&&\multicolumn{1}{c}{(1)}&\multicolumn{1}{c}{(2)}&\multicolumn{1}{c}{(3)}&\multicolumn{1}{c}{(4)}&\multicolumn{1}{c}{(5)}&&\multicolumn{1}{c}{(6)}&\multicolumn{1}{c}{(7)}&\multicolumn{1}{c}{(8)}&\multicolumn{1}{c}{(9)}&\multicolumn{1}{c}{(10)}\\
&&\multicolumn{1}{c}{Bias}&\multicolumn{1}{c}{Std. Dev.}&\multicolumn{1}{c}{RMSE}&\multicolumn{1}{c}{SE/SD}&\multicolumn{1}{c}{p; 95}&&\multicolumn{1}{c}{Bias}&\multicolumn{1}{c}{Std. Dev.}&\multicolumn{1}{c}{RMSE}&\multicolumn{1}{c}{SE/SD}&\multicolumn{1}{c}{p; .95}\\
\midrule
\multicolumn{13}{l}{\bf C. JK-SS2}\\
\multirow{2}{*}{N=25}	& {\tt ltrade}	&	-29.035	&	27.597	&	40.034	&	0.672	&	0.599	&&	4.839	&	26.728	&	27.136	&	0.707	&	0.844	\\
					& {\tt ldist}		&	-54.245	&	80.471	&	96.974	&	0.485	&	0.617	&&	-20.435	&	66.444	&	69.452	&	0.618	&	0.806	\\
[1em]																						
\multirow{2}{*}{N=50}	& {\tt ltrade}	&	-3.259	&	7.836	&	8.479	&	0.963	&	0.932	&&	-0.283	&	11.633	&	11.625	&	0.869	&	0.904	\\
					& {\tt ldist}		&	-6.515	&	25.274	&	26.076	&	0.642	&	0.792	&&	-5.351	&	23.101	&	23.690	&	0.858	&	0.890	\\
[1em]																						
\multirow{2}{*}{N=75}	&{\tt ltrade}	&	-1.158	&	5.079	&	5.204	&	0.924	&	0.920	&&	-1.167	&	7.509	&	7.591	&	0.934	&	0.928	\\
					&{\tt ldist}		&	-2.737	&	15.151	&	15.382	&	0.677	&	0.816	&&	-3.377	&	15.085	&	15.444	&	0.882	&	0.892	\\
[1em]																						
\multirow{2}{*}{N=100}	&{\tt ltrade}	&	-0.457	&	3.575	&	3.601	&	0.952	&	0.922	&&	-1.178	&	5.403	&	5.525	&	0.975	&	0.936	\\
					&{\tt ldist}		&	-1.751	&	11.024	&	11.152	&	0.681	&	0.792	&&	-2.928	&	11.014	&	11.386	&	0.901	&	0.908	\\
[1em]																						
\multirow{2}{*}{N=157}	&{\tt ltrade}	&	-0.119	&	2.158	&	2.159	&	0.972	&	0.952	&&	-1.060	&	2.770	&	2.964	&	0.927	&	0.928	\\
					&{\tt ldist}		&	-0.942	&	6.132	&	6.198	&	0.762	&	0.850	&&	-2.210	&	5.821	&	6.221	&	0.978	&	0.916	\\
\midrule
\multicolumn{13}{l}{\bf D. JK-Double}\\				
\multirow{2}{*}{N=25}	& {\tt ltrade}	&	-33.612	&	52.118	&	61.973	&	0.378	&	0.714	&&	1.119	&	23.181	&	23.185	&	0.815	&	0.876	\\
					& {\tt ldist}		&	-32.292	&	33.637	&	46.604	&	1.188	&	0.930	&&	-8.335	&	37.439	&	38.319	&	1.096	&	0.926	\\
[1em]																						
\multirow{2}{*}{N=50}	& {\tt ltrade}	&	-3.152	&	6.346	&	7.080	&	1.189	&	0.958	&&	-1.927	&	10.933	&	11.091	&	0.924	&	0.910	\\
					& {\tt ldist}		&	-3.203	&	14.739	&	15.068	&	1.101	&	0.954	&&	-2.834	&	17.118	&	17.334	&	1.158	&	0.960	\\
[1em]																						
\multirow{2}{*}{N=75}	&{\tt ltrade}	&	-1.286	&	4.488	&	4.664	&	1.045	&	0.962	&&	-2.012	&	7.182	&	7.451	&	0.977	&	0.936	\\
					&{\tt ldist}		&	-1.120	&	10.270	&	10.321	&	0.999	&	0.954	&&	-2.052	&	11.862	&	12.027	&	1.122	&	0.960	\\
[1em]																						
\multirow{2}{*}{N=100}	&{\tt ltrade}	&	-0.643	&	3.288	&	3.347	&	1.035	&	0.946	&&	-1.814	&	5.073	&	5.383	&	1.039	&	0.932	\\
					&{\tt ldist}		&	-0.821	&	7.781	&	7.816	&	0.965	&	0.938	&&	-2.108	&	8.744	&	8.986	&	1.135	&	0.970	\\
[1em]																						
\multirow{2}{*}{N=157}	&{\tt ltrade}	&	-0.180	&	2.062	&	2.068	&	1.017	&	0.968	&&	-1.692	&	2.607	&	3.105	&	0.985	&	0.876	\\
					&{\tt ldist}		&	-0.257	&	4.423	&	4.427	&	1.056	&	0.960	&&	-1.951	&	4.305	&	4.723	&	1.322	&	0.976	\\
\bottomrule
\hline
\end{tabular}
\begin{tablenotes}
\footnotesize
\item[] Notes: FE denotes uncorrected fixed effect estimator; AN-0 denotes analytical correction with 0 lags; JK-SS2 denotes split jackknife in both dimensions; JK-Double denotes delete-one jackknife for observations with the same index in the cross-section and the time-series.
\end{tablenotes}
\end{threeparttable}
\end{table}
\end{landscape}

\clearpage
\end{document}